\documentclass[useAMS,usenatbib]{mn2e}
\usepackage{graphicx}

\title[Selection effects in observed CV samples]{The influence of selection effects on the observed cataclysmic variable population: modelling and application to the Palomar-Green sample}
\author[M.L. Pretorius, C. Knigge and U. Kolb]{Magaretha L. Pretorius,$^{1}$\thanks{E-mail: mlp@astro.soton.ac.uk (MLP); christian@astro.soton. ac.uk (CK); U.C.Kolb@open.ac.uk (UK)}
Christian Knigge$^{1}$\footnotemark[1] and Ulrich Kolb$^{2}$\footnotemark[1]\\
$^{1}$School of Physics and Astronomy, University of Southampton, Highfield, Southampton SO17 1BJ, United Kingdom\\
$^{2}$Department of Physics and Astronomy, The Open University, Walton Hall, Milton Keynes MK7 6AA, United Kingdom}
\begin{document}


\pagerange{\pageref{firstpage}--\pageref{lastpage}} \pubyear{}

\maketitle

\label{firstpage}

\begin{abstract}
Large differences between the properties of the known sample of cataclysmic variable stars (CVs) and the predictions of the theory of binary star evolution have long been recognised.  However, because all existing CV samples suffer from strong selection effects, observational biases must be taken into account before it is possible to tell whether there is an inconsistency.  In order to address this problem, we have modelled the impact of selection effects on observed CV samples using a Monte Carlo approach.  By simulating the selection criteria of the Palomar-Green (PG) Survey, we show that selection effects cannot reconcile the predictions of standard CV evolution theory with the observed sample.  More generally, we illustrate the effect of the biases that are introduced by magnitude limits, selection cuts in $U-B$, and restrictions in galactic latitude.
\end{abstract}


\section{Introduction}

Cataclysmic variable stars (CVs) are semi-detached binary stars with orbital periods ($P_{orb}$) typically of the order of hours, consisting of a white dwarf primary accreting from a companion that is usually a late-type, approximately main-sequence star.  \citet{bible} gives a comprehensive review of the subject.

An important question in current CV research concerns the secular evolution of these systems.  The evolution of CVs is of interest in its own right (it determines the size and properties of the Galactic CV population), but also because the processes thought to drive it are equally relevant to all close binary systems containing a late-type component.  Binary star evolution is driven by loss of orbital angular momentum ($J$) through gravitational quadrupole radiation, and, in at least some cases, the much more efficient mechanism of magnetic braking (e.g. \citealt{perm1}; \citealt{p17}; \citealt{vz}; \citealt{perm3}).

\begin{figure}
 \includegraphics[width=84mm]{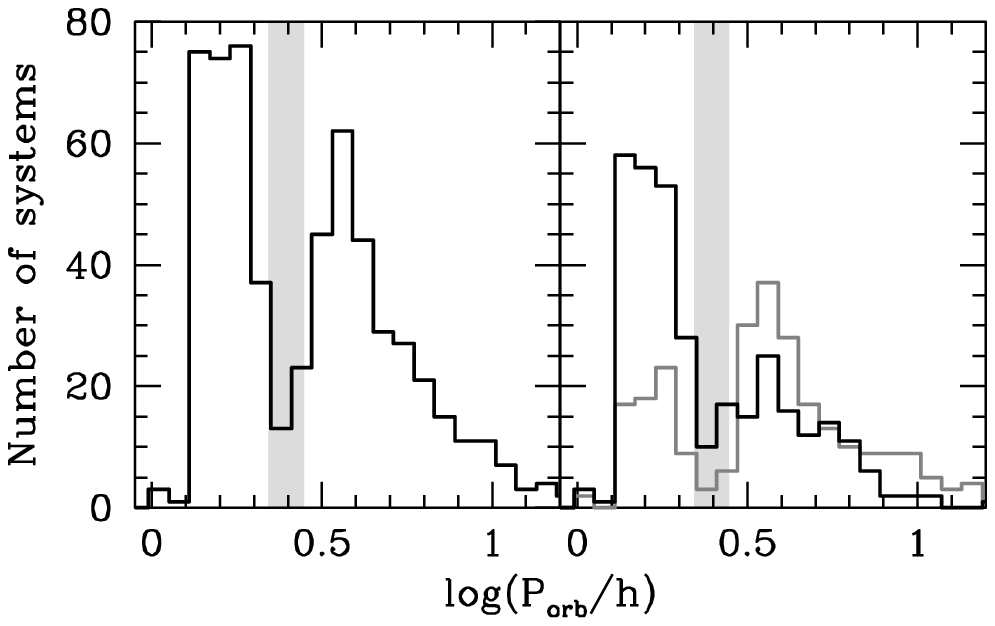}
 \caption{The orbital period distribution of non-magnetic, hydrogen-rich CVs (left hand panel).  Periods are taken from \citet{rkcat}.  The right hand panel shows the same distribution divided into two magnitude bins, $V > 16.5$ (black) and $V \le 16.5$ (dark grey).  The magnitudes are mag1 from \citet{rkcat}, and are actually not in all cases $V$.  We chose $V=16.5$ as a divide because this is roughly the magnitude where it becomes difficult to measure $P_{orb}$ from radial velocity variations with a 2-m class telescope.  The period gap is indicated by light grey shading of the range $2.2\,\mathrm{h} < P_{orb} < 2.8\,\mathrm{h}$ in both panels.
}
 \label{fig:porb}
\end{figure}

The time derivative of the orbital period ($\dot{P}_{orb}$) and orbital angular momentum ($\dot{J}$) are related by
\begin{equation}
\frac{\dot{P}_{orb}}{P_{orb}}=3\frac{\dot{J}}{J}-\frac{2+3q}{1+q}\frac{\dot{M}_1}{M_1}-\frac{3+2q}{1+q}\frac{\dot{M}_2}{M_2}
\label{eq:pdot}
\end{equation}
(e.g. \citealt{bible}), where $M_1$ and $M_2$ are the white dwarf and secondary mass respectively, $\dot{M}_1$ and $\dot{M}_2$ are their time derivatives, and $q=M_2/M_1$ is the mass ratio.  Equation~\ref{eq:pdot} implies that the orbital period distribution of CVs is a useful indicator in the study of their evolution---loss of angular momentum and the resulting mass transfer changes $P_{orb}$.  In practise, $P_{orb}$ is also the only parameter accurately measured for a large number of systems.  Most studies of the evolution of CVs are therefore aimed at accounting for the observed $P_{orb}$ distribution.  This distribution is shown in the left hand panel of Fig.~\ref{fig:porb}. Its most striking features are the period gap (a pronounced drop in the number of systems at $2~\mathrm{h} \la P_{orb}\la 3$~h), and the period minimum (a sharp cut-off in the period distribution of hydrogen rich CVs at about 80~min).  

The period gap is usually explained as resulting from disrupted magnetic braking (e.g. \citealt{perg1}; \citealt{perg2}; \citealt{perg3}).  In a CV approaching the upper edge of the period gap, $M_2$ is approaching the value where the internal structure of a low-mass main-sequence star changes from a deep convective envelope to fully convective.  When the secondary loses its radiative core, its magnetic field structure is proposed to change, causing the magnetic braking mechanism to become less efficient, or to cease operating altogether.  Angular momentum loss then proceeds at the much lower rate allowed by gravitational radiation.  Because of its history of rapid mass loss, the secondary is at this point out of thermal equilibrium, which it now regains by shrinking in radius, thereby losing contact with its Roche lobe, and only regaining it when  $P_{orb}$ has decreased to approximately 2~h.  

The disrupted magnetic braking model is broadly consistent with observational estimates of mass transfer rates ($\dot{M}$) in CVs---these are on average much higher for long-period systems ($P_{orb}\ga 3$~h) than for systems at $P_{orb}\la 3$~h (e.g. \citealt{pat1}).  That magnetic braking should be disrupted has, however, been challenged both on theoretical \citep{t+p92a} and observational \citep{p20} grounds. Nevertheless, disrupted magnetic braking remains the conventional explanation for the period gap, and a cornerstone of CV evolution theory.

The period minimum is a consequence of the response of a secondary of very small mass to continuing mass loss (e.g. \citealt{perm1}; \citealt{p17}; \citealt{perm3}).  The mass loss time-scale increases as $M_2$ decreases, but the thermal time-scale of the secondary increases faster.  When the thermal time-scale exceeds the mass-transfer time-scale, the secondary is not able to shrink fast enough (or in fact expands) in response to mass loss, so that the orbital evolution slowly moves back through longer periods.  CVs in this phase of evolution are referred to as `period bouncers'. 

The rudiments of the theory of CV evolution, outlined above, have long been accepted, yet there are still significant discrepancies between the predictions of theory and the properties of the observed CV sample.  One of the problems is that theoretical models persistently put the period minimum about 10~min short of the observed cut off (see \citealt{kolb02} for a review).  Population synthesis studies also predict that $\simeq 99$\% of the intrinsic galactic CV population should consist of systems with $P_{orb}\la 3$~h, and that about 70\% should be period bouncers (\citealt{pop}; \citealt{howell2}).  The first of these predictions is not reflected in the known sample of CVs---long-period systems comprise significantly more than 1\% of observed CVs (see e.g. the left hand panel of Fig.~\ref{fig:porb}).  Furthermore, only a handful of known systems are likely period bouncers (e.g. \citealt{pat4}).  Clearly, if the standard theory of CV evolution is even approximately correct, the observed CV sample is not representative of the intrinsic population.  However, since different types of CVs differ in intrinsic brightness, and all surveys are flux limited (and the majority severely so) there is no reason to think that an observed sample should be representative of the intrinsic population.  Apparent brightness and large amplitude variability are two factors that obviously affect the discovery probability of a CV.  Since both intrinsic brightness and the frequency of large amplitude brightness variations decrease with $\dot{M}$, short-period, low-$\dot{M}$ systems are necessarily under-represented in the known sample.  The question is only whether these and other selection effects can account for the size of the discrepancy between the period distributions of the predicted and observed samples.  The right-hand panel of Fig.~\ref{fig:porb} goes some way towards illustrating the importance of the apparent magnitudes, by comparing the period distributions of bright and faint sub-samples of known CVs.  The ratio of the number of short- to long-period CVs is larger when only apparantly faint systems are considered.

Although detailed models of the intrinsic CV population exist, very little effort has been made to model even the simplest observational selection effects.  It is therefore not clear to what extent theoretical predictions disagree with observations.  Here we present a Monte Carlo approach that can be used to model any combination of observational selection effects quantitatively, and use it to demonstrate how common selection effects influence the observed CV period distribution.  We then apply our method to a well-defined observational CV sample to show that the intrinsic CV population predicted by the standard theory of CV evolution is indeed inconsistent with observations.

\section{The computational method}
We use a Monte Carlo model to determine the effect of arbitrary selection criteria on observationally constructed CV samples.  The key parameters that describe a CV are $M_1$, $M_2$, $P_{orb}$ and $\dot{M}$.  Given a description for the birthrate of CVs and a recipe for their subsequent evolution, population synthesis methods can be used to generate the predicted intrinsic CV population as a function of these parameters.  We generate a random sample of CVs drawn from such a distribution, and distribute them within a model galaxy.  By modelling the spectral energy distribution (SED) of each CV, we can construct a predicted sample corresponding to an observational CV sample subject to given selection criteria.

\subsection{The predicted intrinsic CV population}
The starting point of our modelling is the population of CVs resulting from model pm5 of \cite{pop}.   This is the present day population of CVs that results from the birth rate model of \cite{dK} when systems are evolved using the magnetic braking prescription of \cite{vz} in the case of secondary stars with radiative cores, and gravitational radiation alone for fully convective secondaries\footnote{\cite{pop} considered several combinations of birth rate models and magnetic braking prescriptions, and found similar resulting populations.  
The strength of magnetic braking is, however, not known with great certainty, and \cite{it03} have proposed a magnetic braking law that is significantly weaker than the \cite{vz} prescription.}.  The population synthesis model provides us with a probability distribution function (PDF) over the parameters $P_{orb}$, $\langle \dot{M} \rangle$ (the secular average $\dot{M}$), $M_1$, and $M_2$.

This model uses a bipolytropic treatment of the secondary, which becomes fully convective when $M_2=0.25\,\mathrm{M_\odot}$.  The resulting period gap is between roughly 2.8 and 3.2~h, which is not where the observed gap is.  This shortcoming can in principle be overcome and does not have an important effect on the results we will derive, since the recalibration needed to improve the match to the observed location of the period gap has no serious repercussions for the overall population synthesis results.  The PDF is zero at $P_{orb}=2.92\,\mathrm{h}$, and we use this value to distinguish between short- and long-period systems in the theoretical distribution.  $M_1$ is kept constant throughout the evolution; i.e., all the material transferred from the secondary to the primary is assumed to be ejected during nova eruptions.  Period bounce occurs at  $M_2 \simeq 0.07\,\mathrm{M_\odot}$\footnote{We will simply define the period bouncers in our population as systems with $M_2 \le 0.07\,\mathrm{M_\odot}$.  This is close to correct, although systems actually bounce at a small range of $M_2$.}, placing the period minimum at about 64~min.  In this population, 99.7\% of all systems are below the period gap, and 72\% of all systems are period bouncers.  

Magnetic systems are not accounted for separately, so if they evolve differently, as has been proposed at least for polars (e.g. \citealt{ww93}; \citealt{lww94}; \citealt{ww02}), they are not included in the population synthesis model.  Furthermore, the largest secondary mass considered was $1\,\mathrm{M_\odot}$, and all secondaries are assumed to be completely unevolved at the onset of mass transfer.  

It is worth pointing out that available population synthesis models assume a galactic age of 10~Gy or more.  There is evidence that the galactic thin disc is only 7 to 8~Gy old, while the thick disc is no older than 10~Gy, with a scale height of $\sim$750~pc, and a density more than a factor of 10 lower than that of the thin disc (e.g. \citealt{thick}; \citealt{siegel02}; \citealt{reid05}).  If this is correct, the oldest CVs in the predicted population should be removed, which would reduce the predicted number of period bouncers.

\subsection{The model galaxy}
Our model of the Galaxy consists solely of an axisymmetric disk, with no spiral structure, no additional thick disc, no halo and no galactic bulge.  This is consistent with the Population I assumption made in our population synthesis model and is reasonable, since the bulk of the intrinsic CV population resides in the (thin) galactic disk (e.g. \citealt{popii}).

Every system is assigned a position $(r,z,\phi)$ in space, where $z$ is height above the galactic plane, $r$ is distance from the galactic centre, and $\phi$ is azimuth.  The positions are generated randomly in such a way that the overall distribution of CVs produces exponential density profiles:
\begin{equation}
\rho(z) \propto \exp (-|z|/H_z)\mathrm{, and}
\label{eq:3exp}
\end{equation}
\begin{equation}
\rho(r) \propto \exp (-r/H_r),
\end{equation}
where the radial scale length is $H_r=3\,000$~pc, and the vertical scale height is 
\begin{equation}
H_z = \cases{120\,\mathrm{pc}  &for long-$P_{orb}$ systems\cr
260\,\mathrm{pc}  &for normal short-$P_{orb}$ systems\cr
450\,\mathrm{pc}  &for period bouncers.\cr}
\label{eq:3discs}
\end{equation}
Adopting different scale heights for these three sub-populations is motivated by the fact that the long-period, short-period (pre-period bounce), and period bounce regimes are successive stages in the evolution of a single CV.  The age of a CV is the sum of the time spent as a detached binary (pre-CV) and the time since becoming semi-detached, and systems first make Roche-lobe contact at a range of orbital periods.  This means that there is a spread in the ages of CVs at fixed period (besides that caused by period bounce), and, in fact, some overlap between the ages of sub-populations (see e.g. \citealt{ks96}).  Nevertheless, sub-populations have different representative, average ages.  Long period CVs are younger than $\sim10^{8.4}$~y, while short-period, pre-period bounce systems are older than $\sim10^{8.8}$~y, and it takes $\sim10^{9.4}$~y to reach the period minimum (e.g. \citealt{howell}).  The scale heights of the 3 sub-populations given in equation~\ref{eq:3discs} were chosen to be representative of stellar populations with ages of $\simeq 10^{8.4}$~y, $\simeq 10^{9.2}$~y, and $\simeq 10^{9.6}$~y \citep{rc86}. 

We assume a galactic centre distance of $7\,620\,\mathrm{pc}$ \citep{gcd}, and ignore the small offset of the Sun from the galactic plane; i.e., the Earth is placed at $(r,z,\phi)=(7\,620\,\mathrm{pc},0,0)$.  For computational simplicity, we divide the galaxy along the two planes of symmetry that include both the earth and the galactic centre (the planes $z=0$ and $\phi=0$) and simulate only one quadrant of the galaxy; i.e. $z \ge 0$ and $0 \le \phi < \pi$.

Interstellar extinction is computed by integrating the density of the interstellar medium 
\begin{equation}
\rho_{ISM} \propto \exp \left ( \frac{7\,620\,\mathrm{pc}-r}{4\,500\,\mathrm{pc}} \right ) \times \exp \left (\frac{-|z|}{140\,\mathrm{pc}} \right )
\label{eq:ism}
\end{equation}
\citep{ism} along the line of sight to each system to obtain the neutral hydrogen column density, $N_H$.  
We find the proportionality constant in equation~\ref{eq:ism} by setting $A_V=30$ for an object at the galactic centre.  Extinctions in the $U$ and $B$-bands are taken as $A_U=1.531A_V$ and $A_B=1.324A_V$ \citep{extinc}.

By equation~\ref{eq:3discs}, the Gaussian scale height of 190~pc found empirically by \cite{pat1} is compatible with the vertical distribution we use for the youngest systems.  This measurement of the scale height was based on the known sample, which is strongly biased towards intrinsically bright systems (this bias was even more pronounced two decades ago), and it seems reasonable that it may reflect the galactic distribution of only the brighter, younger systems.  Although younger stellar populations are expected to be more concentrated towards galactic mid-plane than older populations, the observational evidence in the case of CVs is slim or nonexistent (\citealt{sh92}; \citealt{vanp}).  As a check, we therefore also performed simulations using the same vertical density profile for all systems.  In those simulations we used 
\begin{equation}
\rho(z) \propto 0.80\,\mathrm{sech}^{2}(z/323\,\mathrm{pc})+0.20 \exp (-|z|/656\,\mathrm{pc}),
\label{eq:gauld}
\end{equation} 
which results from counts of M stars \citep{galaxy}, or a Gaussian profile, 
\begin{equation}
\rho(z) \propto \exp \left [- \left (|z|/H_z \right )^2 \right ],
\label{eq:pat}
\end{equation}
with scale height $H_z=190$~pc \citep{pat1}.

\subsection{The spectral energy distribution of CVs}
Four components, namely the accretion disc, bright spot, white dwarf, and secondary star, are included in the model of the overall SED for each CV in our simulations.  

We used a slight modification of the code described in \cite{disc} to compute a 4D grid of accretion disc models over the variables $\dot{M}_d$ (the rate at which mass flows through the disc), $M_1$, the orbital inclination $i$, and the outer disc radius $r_d$.  We set the viscosity parameter $\alpha=0.01$ for $\dot{M}_d \le 10^{13}\,\mathrm{g/s}$, and $\alpha=0.8$ otherwise\footnote{This choice was made mostly for computational convenience and is not very physical, but the influence of $\alpha$ is small in all regards except $U-B$ colours \citep{disc}.  This is discussed further in Section 4.}.  The disc model assumes a steady state disc structure and produces a blackbody spectrum for optically thick regions, while for a vertical optical depth $\la 1$, the same procedure as explained in \cite{disc} is adopted, so that emission lines appear in these regions.  As will be explained in the next section, we take $\dot{M}_d=\langle \dot{M} \rangle$ only for systems with stable discs (these are a small minority of the population).  

CVs are assumed to be randomly inclined with respect to Earth, so we select $\cos i$ for each system as a uniformly distributed random number between 0 and 1.  The inner disc radius is set equal to the radius of the primary ($R_1$), which is calculated using the approximate white dwarf mass-radius relation of \cite{massr}.  The calculation of $r_d$ will be explained in the next section.

The bright spot is modelled as a blackbody extending over an azimuth of $10^\circ$ on the outer disc rim (see e.g. \citealt{spot}).

The flux contribution of the secondary is computed using the spectral type --$P_{orb}$ relation of \cite{sd}:
\begin{equation}
Sp(2) = \cases{26.5 - 0.7 P_{orb}/\mathrm{h} &for $P_{orb} < 4\,\mathrm{h}$\cr
33.2 - 2.5 P_{orb}/\mathrm{h} &for $P_{orb} \ge 4\,\mathrm{h}$, \cr}
\label{eq:sd}
\end{equation}
where $Sp(2)=0$ means a spectral type of G0, $Sp(2)=10$ is K0, $Sp(2)=20$ is M0, etc\footnote{Because the predicted period minimum is shorter than observed, we apply equation~\ref{eq:sd} beyond the $P_{orb}$ range where it can be expected to be valid.  Since these very low-mass secondaries are faint compared to other components of the binaries, this is not an important concern.  The $M_{Vmax}$--$P_{orb}$ relation discussed below is also extrapolated to periods below the range where it is known to apply.}.  The secondary is assigned the absolute magnitudes that a main-sequence star of the same spectral type would have.  This is not strictly correct, since CV secondaries are not in thermal equilibrium, but unevolved secondaries are not expected ever to be large contributors to the optical flux.  We may nevertheless improve our treatment of secondary stars in the future by using the sequence of \cite{ck06}.  In systems with $M_2 \le 0.07\,\mathrm{M_\odot}$ (i.e. period bouncers), the flux contribution of the secondary is neglected.  

We estimate the effective temperature ($T_{eff}$) of the primary using 
$$T_{eff} = 15\,600 \,\mathrm{K}\left(\frac{\langle \dot{M} \rangle}{10^{16}\,\mathrm{g\,s^{-1}}}\right)^{1/4} \left(\frac{R_1}{10^{9}\,\mathrm{cm}} \right)^{-1/2}$$
\citep{tb}.  The absolute $U$, $B$, and $V$ magnitudes of the primary are then found using the tabulation for pure H white dwarfs of \cite{wd}.   In systems with $\dot{M}_d>2.0 \times 10^{17}$~g/s and $i>86^\circ$, we assume that the white dwarf is completely shielded by the disc, and makes no contribution to the flux\footnote{In detail, the issue of shielding by the disc is more complicated than we have allowed for; see e.g. \cite{smak92}, \cite{mmh82}, and Knigge et al. (2000, 2004).}.

\subsection{Dwarf nova outburst cycles}
Most CVs have values of $\dot{M}_2$ that cause their discs to be subject to a thermal instability; these systems are the dwarf novae (we will abbreviate dwarf nova(e) to DN), and they spend most of their time with  $\dot{M}_d \ne -\dot{M}_2$, so that it is incorrect to use $\langle \dot{M} \rangle$ in the disc luminosity calculation.  DN are characterised by outbursts during which their luminosity increases by typically 2--5 mag.  Outburst durations are days to tens of days, and the recurrence interval ranges from tens of days to decades.  

To determine whether a system is a DN, we use equations 38 and 39 of \cite{dim}; systems with 
\begin{displaymath}
\langle \dot{M} \rangle > 9.5 \times 10^{15} \alpha^{0.01}\left(\frac{M_1}{\mathrm{M_\odot}}\right)^{-0.89}\left(\frac{0.7R_L}{10^{10}\,\mathrm{cm}}\right)^{2.68}\,\mathrm{g/s}
\end{displaymath}
or 
\begin{displaymath}
\langle \dot{M} \rangle  < 4.0 \times 10^{15} \alpha^{-0.004}\left(\frac{M_1}{\mathrm{M_\odot}}\right)^{-0.88}\left(\frac{R_1}{10^{10}\,\mathrm{cm}}\right)^{2.65}\,\mathrm{g/s}
\end{displaymath}
have stable discs.  None of the CVs in the theoretical intrinsic population satisfy the first inequality, so that we have no systems that are stable on the lower branch of the S-curve (see \citealt{dim}).  All short-period systems in our model population are DN, and less than 0.1\% of CVs in this population are permanently bright, nova-like (NL) systems.

In order to calculate the disc spectrum correctly for the systems with unstable discs, we need to find the mass accretion rate through the disc in outburst ($\dot{M}_{dO}$) and quiescence ($\dot{M}_{dQ}$).  These are related to the average value by
$$\langle \dot{M} \rangle = C \dot{M}_{dO} + (1-C)\dot{M}_{dQ},$$
where $C$ is the outburst duty cycle (we assume that outbursts are simple top-hat functions).  For DN that have not yet reached the period minimum, we use the $M_{Vmax}$--$P_{orb}$ relation of \cite{harrison} to predict the absolute $V$ magnitude at maximum, but since the candidate period bouncers reach brighter maxima than given by this relation (these systems show only superoutbursts), we use
\begin{equation}
M_{Vmax} = \cases{5.92-0.383P_{orb}/\mathrm{h} &for normal DN\cr
5 &for period bouncers. \cr}
\label{eq:mv}
\end{equation}
We assume that all of the outburst light originates from the disc, and use a bolometric correction of $-1.8$ (following \citealt{pat02}) to find the disc luminosity ($L_d$) and hence the mass accretion rate through the disc at maximum,
\begin{equation}
\dot{M}_{dO}=\frac{2R_1L_d}{GM_1}\,.
\label{eq:dotmo}
\end{equation}
The rate at which mass flows through the disc in quiescence is then 
$$\dot{M}_{dQ}=\frac{\langle \dot{M} \rangle - C \dot{M}_{dO}}{1-C}\,.$$
No clear empirical relation exists between the duty cycle and other observables, but for normal pre-period bounce DN a duty cycle of $\sim$10\% is in accord with observations (e.g. \citealt{bible}; \citealt{ak})\footnote{Most (possibly all) short-period DN are SU UMa stars and have superoutbursts in addition to normal outbursts.  However, since the latter are considerably more frequent, we ignore this complication.}.  For period bouncers, a duty cycle of $\sim 5 \times 10^{-3}$ is indicated by observations of the best candidates\footnote{The duration and recurrence time of outbursts in these systems are uncertain.  WZ Sge, EG Cnc, and AL Com are considered good period bouncer candidates (e.g. \citealt{p98}; \citealt{pat4}), and appear to have outburst duty cycles of about 0.5\%, but the observational baseline is not long enough for this to be firmly established.}.

\begin{table}
 \centering
 \caption{The DN outburst duty cycles we adopt.  The final column lists the fraction of the total intrinsic population represented by DN with each duty cycle.}
 \label{tab:outbursts}
 \begin{tabular}{@{}lll@{}}
 \hline
                & DN outburst duty    & Percentage of the \\
                & duty cycle          & population\\
 \hline
Period bouncers & $5 \times 10^{-3}$  & 8.2  \\
                & $5 \times 10^{-4}$  & 44   \\
                & 0                   & 20   \\
Normal DN       & 0.1                 & 1.1  \\
                & 0.01                & 22   \\
                & $5 \times 10^{-3}$  & 2.1  \\
                & $5 \times 10^{-4}$  & 2.6  \\
 \hline
 \end{tabular}
\end{table}

As it turns out, however, equation~\ref{eq:mv} and the mass transfer rates given by evolution theory cannot be reconciled with $C=0.1$ ($C=5 \times 10^{-3}$) for the majority of normal DN (period bouncers).  Duty cycles as large as these would imply  $\langle \dot{M} \rangle < C \dot{M}_{dO}$ and thus $\dot{M}_{dQ} < 0$.  This is because equation~\ref{eq:dotmo} gives $\dot{M}_{dO}$ between $\sim 2 \times 10^{16}\,\mathrm{g/s}$ and $\sim 8 \times 10^{17}\,\mathrm{g/s}$, while $\langle \dot{M} \rangle$ extends as low as $5 \times 10^{13}\,\mathrm{g/s}$.  We deal with this problem by assuming that typical DN duty cycles are lower than those suggested by known systems. This is plausible, since the known sample of DN is likely to be biased towards systems with unusually frequent outbursts.  Our grid of accretion disc models in fact extends only to $\dot{M}_d= 3.2 \times 10^{12}\,\mathrm{g/s}$ (by this point the disc is in any case always fainter than the white dwarf).  Therefore we require
$$\dot{M}_{dQ}=\frac{\langle \dot{M} \rangle - C \dot{M}_{dO}}{1-C} \ge 3.2 \times 10^{12}\,\,\mathrm{g/s.}$$
For normal DN, if $C=0.1$ does not satisfy the above inequality, we try successively $C=0.01$, $C=5 \times 10^{-3}$, and $C=5 \times 10^{-4}$, and adopt the first value that gives $\dot{M}_{dQ} \ge 3.2 \times 10^{12}\,\,\mathrm{g/s}$.  For period bouncers, we start with $C=5 \times 10^{-3}$, and use $C=5 \times 10^{-4}$ if needed.  If $C=5 \times 10^{-4}$ also leads to $\dot{M}_d < 3.2 \times 10^{12}\,\mathrm{g/s}$, we assume that the system will never be observed in outburst, and that the disc luminosity is negligible in quiescence.  Table~\ref{tab:outbursts} summarises the outburst duty cycles used.  If we require only $\dot{M}_{dQ} \ge 0$, then DN that must have $C < 5 \times 10^{-4}$ comprise 19\% of all systems, and less than 1\% of the total population consists of DN that can have $C \ge 0.1$.   

Of course the procedure described above is very contrived, but, given that it is not possible to reproduce the DN outburst properties expected on empirical grounds, this approach is at least reasonable.  Most known DN were discovered through their outbursts.  Assuming that 
(i) the space density of CVs is $5 \times 10^{-5}\,\mathrm{pc^{-3}}$; 
(ii) all DN reach $M_V=5$ in outburst; 
(iii) $C=5 \times 10^{-4}$ implies an outburst recurrence time of 300~y (i.e. that $1/6$ of these systems had an outburst during the last 50~y); and 
(iv) every DN that reached $m_V=11$ at maximum during the last 50~y was discovered and recognised as a DN, 
our treatment of outburst behaviour implies that we should know about 10 systems with $C=5 \times 10^{-4}$.  Given the uncertainty over space density, the practically unquantifiable nature of the probability of discovering a DN in outburst, and the fact that we do know several DN that have had only one observed outburst (GW Lib being the best known example), we do not consider the---admittedly surprising---duty cycles to be in themselves inconsistent with observations.  \cite{p98} points out that a large population of low-$\dot{M}$ CVs could remain undiscovered if these systems have very long outburst recurrence times, and that, while this is not impossible, there is no empirical evidence of such behaviour.  We can add that (accepting the mass transfer rates that result from gravitational radiation alone, that DN outbursts are caused by an increase in $\dot{M}_d$, and that DN reach $M_V \simeq 5$ at maximum) very long outburst recurrence times are not only possible, but indeed inevitable.

We emphasise that equation~\ref{eq:mv} is only used to estimate $\dot{M}_{dO}$ and $\dot{M}_{dQ}$, which are then used to calculate absolute magnitudes from our 4D grid of accretion disc models.  We assume that the probability of catching a DN in outburst is equal to the duty cycle (which is correct for a single-epoch survey), and again use a Monte-Carlo method to simulate some systems in outburst and others in quiescence.  The outer disc radius is set to $r_d=0.7 R_L$ for NLs and DN in outburst, and $r_d=0.5 R_L$ for quiescent systems \citep{haw}, where $R_L$ is the Roche-lobe radius of the primary.

Large scatter in observationally inferred $\dot{M}$ at fixed period (\citealt{pat1}; \citealt{brian87}), and---more compellingly---the coexistence of different subtypes of CVs in the same period intervals, imply that $\dot{M}$ also fluctuates around the secular mean on time-scales that are too long to be observed, but shorter than the binary evolution time-scale (see e.g. \citealt{hkl89}; \citealt{www}; \citealt{kfkr95}; \citealt{mf98}; \citealt{br04}; \citealt{rzk00}).  The fact that these cycles are not included in the population synthesis model used here explains why our population contains no short-period NL variables.

In addition to the treatment of DN outburst cycles described above, we will explore two limiting cases, namely (i) $\dot{M}_d=\langle \dot{M} \rangle$ (which implies maximising $\dot{M}_{dQ}$ and setting $C=0$), and (ii) $\dot{M}_{dQ}=0$ (which is equivalent to adopting the highest allowed value of $C$).

\subsection{Summary of computational details}

\begin{table}
 \centering
 \caption{Summary of the models considered here.}
 \label{tab:models}
 \begin{tabular}{@{}lll@{}}
 \hline
Galactic structure &  DN outburst properties             & Model \\
 \hline
Equation \ref{eq:3exp} and \ref{eq:3discs} & as described in 2.4                 & A1 \\
Equation \ref{eq:gauld}                    & as described in 2.4                 & A2 \\
Equation \ref{eq:pat}                      & as described in 2.4                 & A3 \\
Equation \ref{eq:3exp} and \ref{eq:3discs} & $\dot{M}_d=\langle \dot{M} \rangle$ & B1 \\
Equation \ref{eq:gauld}                    & $\dot{M}_d=\langle \dot{M} \rangle$ & B2 \\
Equation \ref{eq:pat}                      & $\dot{M}_d=\langle \dot{M} \rangle$ & B3 \\
Equation \ref{eq:3exp} and \ref{eq:3discs} & $\dot{M}_{dQ}=0$                    & C1 \\
 \hline
 \end{tabular}
\end{table}

\begin{figure*}
 \includegraphics[width=168mm]{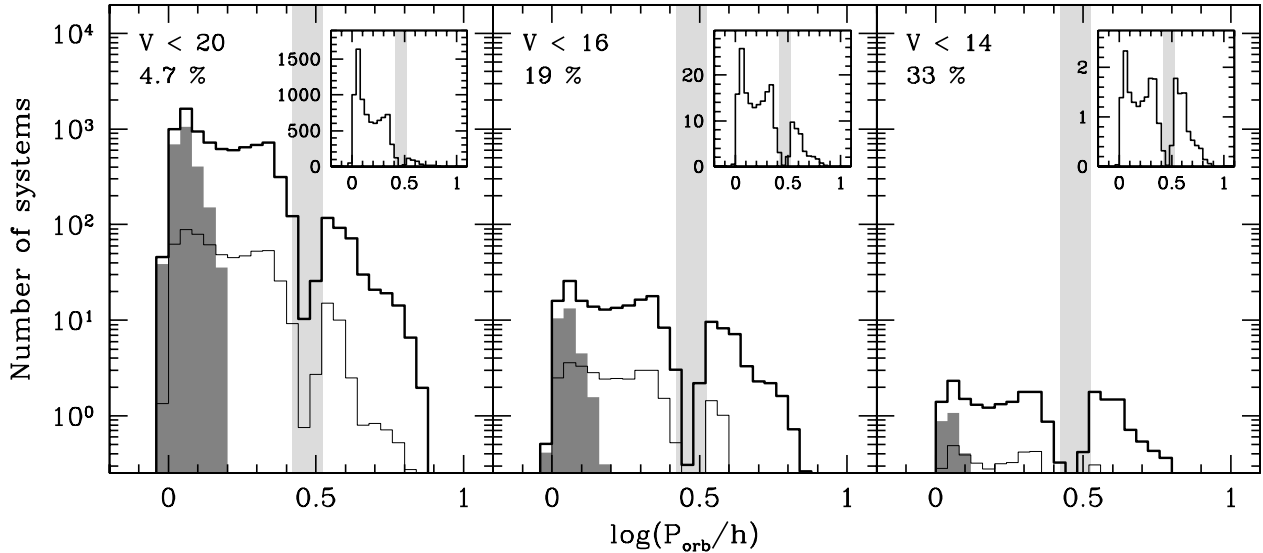}
 \caption{Magnitude limited samples produced by model A1 (see Table~\ref{tab:models}).  The magnitude limit is given in each panel, together with the fraction of the sample with periods above the period gap.  Here, and in all other predicted period distributions we will show, light grey shading of the range $2.63\,\mathrm{h} < P_{orb} < 3.35\,\mathrm{h}$ indicates the model period gap (see Section~2.1).  Bold histograms are all systems detected; the contribution of period bouncers is shaded in dark grey; the fine black histograms show DN that were simulated in outburst.  The ratio of short- to long-period systems is 95:5, 81:19, and 67:33 at magnitude limits of $V<20$, $V<16$, and $V<14$ respectively.  Period distributions of all systems detected at each magnitude limit are plotted with linear vertical scales in the insets.
}
 \label{fig:flim}
\end{figure*}

In general, we find that the accretion disc is the brightest SED component in high-$\dot{M}$ CVs (i.e. DN in outburst and NL systems).  The white dwarf dominates the optical flux in a significant fraction of DN in quiescence, both normal systems and period bouncers.  The secondary star does not dominate the optical flux of any CV in our model.

In order to allow for uncertainties in assigning appropriate galactic scale heights and outburst duty cycles to our CVs, we have run a variety of models.  All the models that we consider here are listed in Table~\ref{tab:models}.  We view model A1 to be the most realistic amongst these, and will focus on this model in the next section, but we also show results from other combinations of model galaxies and DN outburst properties for the sake of comparison (Section 4). 

We note again that our population synthesis model does not include evolved systems, and does not explicitly include magnetic CVs.  Furthermore, our SED modelling does not make allowance for discless systems such as polars.  These restrictions should be kept in mind when interpreting the results presented in the next section.

\section{Selection effects}
In the present study, we restrict our attention to three of the most common selection effects present in observational CV samples, namely magnitude limits, selection for blue $U-B$, and restricted ranges in galactic latitude.  The impact of large amplitude variability (DN outbursts and nova eruptions), emission lines, and X-ray emission on discovery probability will be discussed elsewhere. 

\subsection{Optical flux limits}
Bias towards finding apparently bright objects is the simplest selection effect, and a very important one.  The properties of (bolometric) magnitude limited samples have been studied before, with the assumptions that the observed radiation is the time-averaged accretion luminosity, and that CVs are distributed uniformly in space (\citealt{pop}; \citealt{howell}).  These assumptions make a comparison with data almost meaningless (as was recognised by \citealt{pop}).  Magnitude limits together with galactic structure were considered by \cite{rit1}, \cite{rit2}, and \cite{rit3}.  However, the intrinsic CV populations assumed in these studies were not based on full population synthesis models, only bright CVs (NLs and DN in outburst) were taken into account, and DN outbursts were not treated satisfactorily (it was assumed that all DN will be observed in outburst, and that $\dot{M}_1$ exceeds $\dot{M}_2$ by a constant factor for all systems in outburst).

Fig.~\ref{fig:flim} displays period histograms of samples produced by our standard model A1, with three different magnitude limits ($V<20$, $V<16$, and $V<14$).  In this figure (and, with the exception of Fig.~\ref{fig:comparison}, all other model period histograms we will show) the number of systems in each bin is scaled to reproduce a local space density of $5 \times 10^{-5}\,\mathrm{pc^{-3}}$ (where local space density means counting systems inside a radius of 100 pc)\footnote{The CV space density is not known even to within a factor of 10 (see e.g. \citealt{rit1}; \citealt{shara}; \citealt{politano}; \citealt{p98}; \citealt{schwope02}; \citealt{thors}).  Our choice of $5 \times 10^{-5}\,\mathrm{pc^{-3}}$ lies within the range suggested by different studies.  When the model population is normalised to the formation rate of single white dwarfs, the predicted mid-plane CV space density is $1.8 \times 10^{-4}\,\mathrm{pc^{-3}}$.}.  The shape of the period distribution depends strongly on the magnitude limit.  As is expected from the fact that long-period CVs are intrinsically brighter, the fraction of the total sample made up by long-period systems increases for brighter magnitude limits; it is 4.7\%, 19\%, and 33\% for $V<20$, 16, and 14 respectively.  Dark grey shaded histograms represent period bouncers in each panel of Fig.~\ref{fig:flim}, while the contributions of DN simulated in outburst are shown by the fine black histograms.

\begin{figure}
 \includegraphics[width=84mm]{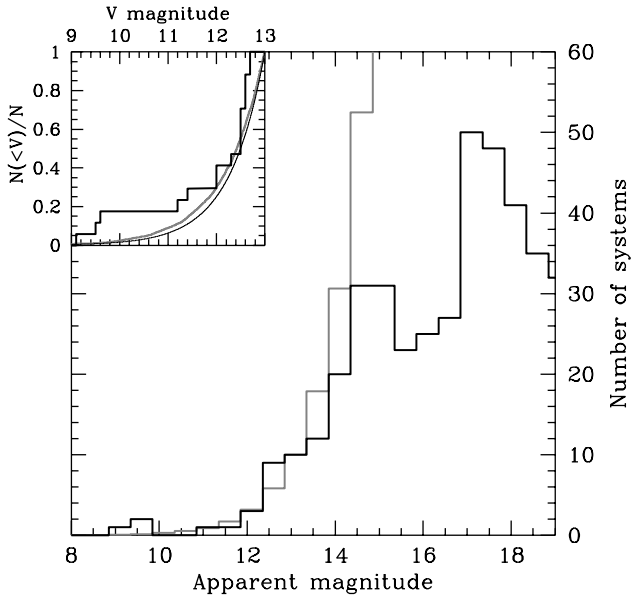}
 \caption{The distribution in apparent magnitude of known CVs (non-magnetic, hydrogen-rich systems in the \citealt{rkcat} sample; black), and a complete sample produced by model A1 (grey; this histogram is arbitrarily scaled).  Systems with $P_{orb} \ge 5\,\mathrm{h}$ were omitted form both samples (see text).  The inset shows cumulative distributions for known CVs (black) and the model sample (grey), together with a sample of objects that are uniformly distributed in space and all have the same absolute magnitude (fine black curve).  Magnitudes used for the histogram of the known sample are mag1 from \citealt{rkcat}, and are not in all cases $V$-band.
}
 \label{fig:mags}
\end{figure}

\begin{figure}
 \includegraphics[width=84mm]{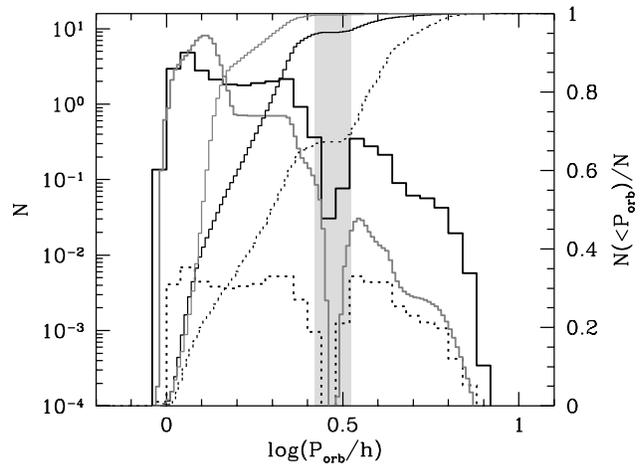}
 \caption{Orbital period histograms and cumulative histograms of the intrinsic population (dark grey) and samples with magnitude limits of $V=20$ (solid black) and $V=14$ (dotted).  The intrinsic and $V<20$ distributions are both normalised, while the $V<14$ histogram is scaled by the normalisation factor of the $V<20$ distribution.  The spike near the period minimum is increasingly suppressed at brighter magnitude limits.  The broad feature at $0 \la \log(P_{orb}/\mathrm{h}) \la 0.2$ is strongly suppressed in the magnitude limited samples.
}
 \label{fig:comparison}
\end{figure}

Of course the known CV sample is not complete to 20th or even 16th magnitude, which raises the question of whether there is a magnitude limit for which we can compare known CVs directly to our predicted magnitude limited samples.  Fig.~\ref{fig:mags} shows the apparent magnitude distribution of known non-magnetic, hydrogen-rich CVs from \cite{rkcat}, with the distribution in $V$ of the complete model sample over-plotted in grey.  Systems with $P_{orb} \ge 5\,\mathrm{h}$ are omitted from both the observed and model populations, because CVs with evolved secondaries dominate the population at periods above $\simeq5$~h (\citealt{bbkw98}; \citealt{bk00}; \citealt{phr03}).  The known samples with $P_{orb} < 5\,\mathrm{h}$ and $V<12$ and $V<13$ contain only 5 and 17 systems respectively.  A Kolmogorov-Smirnov (KS) test shows that the magnitude distribution of known CVs becomes marginally consistent with that of the model sample only at $V\la13$.  The probability that the model and observed distributions are drawn from the same parent population is 0.08, 0.004, and $6 \times 10^{-16}$ for limiting magnitudes of $V<13$, $V<14$ and $V<16$ respectively.  This result is somewhat model-dependent.  The inset in Fig.~\ref{fig:mags} shows cumulative distributions for the observed (bold black) and model (dark grey) samples.  For comparison, the fine black curve represents objects with identical intrinsic brightness, that are uniformly distributed in space.  Since the known sample is likely to be incomplete even for $V<13$, it should not be compared to the predicted magnitude limited samples presented here.

The absence of a spike at the period minimum in the observed distribution has been the subject of several theoretical papers.  \cite{bk02} and \cite{ksh02} aim to explain this by removing the spike from the intrinsic distribution, whereas \cite{kb99} argue that it could possibly be a selection effect.  Our results show that, relative to the whole population detected at a given magnitude limit, the spike at the period minimum indeed decreases in prominence at brighter magnitude limits, although it is present even in the sample with $V<14$.  Fig.~\ref{fig:comparison} illustrates that the spike cannot be discerned in the intrinsic period distribution.  The broader feature between roughly $\log(P_{orb}/\mathrm{h})=0$ and 0.2, caused by period bounce as well as the increasing evolutionary time scale of pre-period bounce systems, is very strongly suppressed, even at $V<20$.

\subsection{Blue optical colours}
\begin{figure}
 \includegraphics[width=84mm]{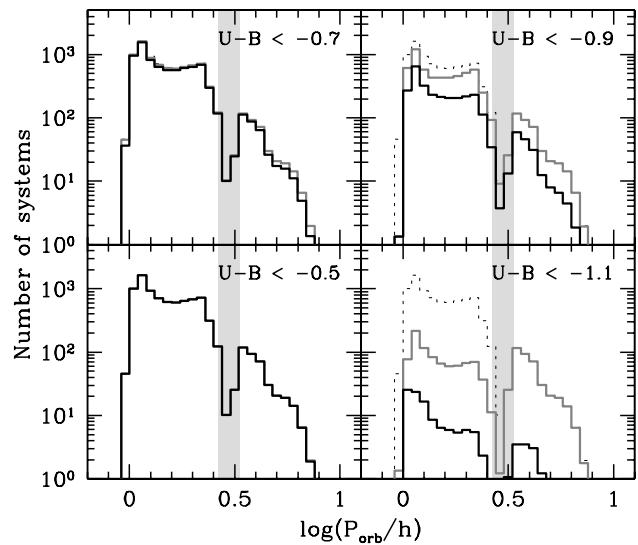}
 \caption{Period histograms of systems with $V<20$ (dotted), and systems with $V<20$ and the $U-B$ indicated in each panel (black).  To illustrate the effect of reddening, dark grey histograms show the distributions obtained when the same blue cuts are applied to the intrinsic colours.  Almost all systems with $V<20$ also have both $(U-B)_0<-0.7$ and $U-B<-0.7$, and most long-period CVs have $(U-B)_0<-1.1$.  These samples were produced by model A1.
}
 \label{fig:blue}
\end{figure}

Most known CVs have $(U-B)_0<-0.5$ (e.g. \citealt{colours}).  This is because high-$\dot{M}$ systems have high continuum colour temperatures, while known low-$\dot{M}$ systems have the Balmer discontinuity in emission.  However, in the systems with lowest $\dot{M}$, the (accretion heated) white dwarf is expected to dominate the optical flux, and this white dwarf may be quite cool ($T_{eff} \la 10\,000$~K).  Selection cuts in $U-B$ may therefore exclude the faintest CVs.

Two $UV$-excess surveys covering large areas at high galactic latitude have produced reasonably large CV samples.  These are the Palomar-Green (PG) Survey ($B<16.2$; $U-B<-0.46$; \citealt{pg}), and the Edinburgh-Cape Blue Object (EC) Survey ($B<16.5$; $U-B<-0.4$; \citealt{ec}).  Follow-up of the EC Survey is still far from complete; the PG Survey will be considered in more detail in Section~\ref{sec:pg}.  

\begin{figure}
 \includegraphics[width=84mm]{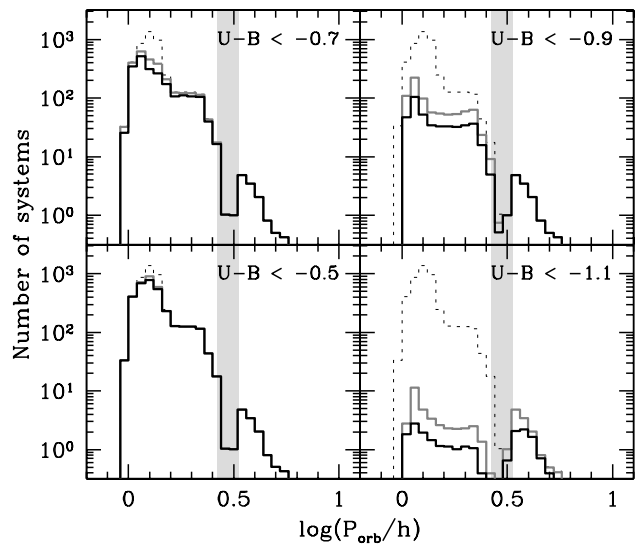}
 \caption{Period histograms of samples with the same $U-B$ and $(U-B)_0$ cuts as in Fig.~\ref{fig:blue}, but here applied to a volume limited sample.  Dotted histograms show all systems at $d<400\,\mathrm{pc}$; black are samples with observed colours satisfying the blue cut indicated in each panel, and grey shows systems with intrinsic colours blue enough to satisfy the selection cuts.  These samples were produced by model A3; i.e., the volume limited sample is representative of the intrinsic population.
}
 \label{fig:blue_v}
\end{figure}

Fig.~\ref{fig:blue} shows the effect of a magnitude limit together with different $U-B$ selection criteria.  In every panel the dotted histogram shows all systems with $V<20$, and the solid black histogram shows systems that also satisfy the blue cut.  It is only at $U-B<-0.9$ that the blue selection starts removing a large fraction of systems from the magnitude limited sample.  

In Fig.~\ref{fig:blue_v} we display $P_{orb}$ distributions resulting from blue cuts imposed on a volume limited sample.  It is seen that even $U-B<-0.5$ excludes systems near the period minimum, while $U-B<-0.7$ introduces a severe bias against short-period systems.  This shows that many CVs in the intrinsic population are not bluer than $U-B=-0.7$, but, in the samples shown in Fig.~\ref{fig:blue}, these systems were already excluded by the magnitude limit.  Thus, $UV$-excess surveys that select objects for $U-B \la -0.7$ are expected to be seriously biased against short-period, low-$\dot{M}$ CVs, but such a blue cut introduces hardly any additional bias in a survey that is also severely flux limited.

The grey histograms in Fig.~\ref{fig:blue} and \ref{fig:blue_v} display samples with the same blue cuts as shown by solid black histograms, but here the blue cuts are applied to the intrinsic colours to illustrate the influence of reddening.  Notice that practically all long-period systems have $(U-B)_0<-0.9$, but some get excluded from observed samples because they are on average more distant than short-period systems, and thus more highly reddened.  This is amplified by the fact that they are (in model A1) also more concentrated towards the galactic plane.

\subsection{Restrictions in galactic latitude}
\begin{figure}
 \includegraphics[width=84mm]{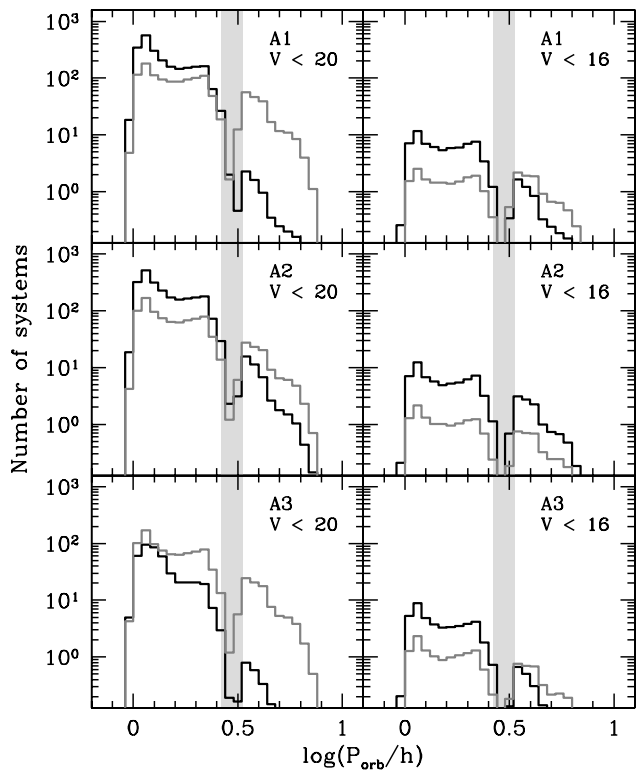}
 \caption{The orbital period distributions of high galactic latitude ($|b|>30^\circ$; black) and galactic plane ($|b|<5^\circ$; dark grey) samples with magnitude limits $V<20$ (left) and $V<16$ (right).  The plots are repeated for all our models of galactic structure.  Surveys at different galactic latitude include very different populations, and, as expected, the effect is larger for deeper samples.
}
 \label{fig:gb}
\end{figure}

Surveys at high galactic latitude hold some promise of uncovering samples that are volume limited.  In a galactic plane survey, on the other hand, intrinsically bright objects will be detected out to larger distances than fainter systems.  The direction of this effect is obvious; we simply show its magnitude in Fig.~\ref{fig:gb}, with period histograms of high (black) and low (dark grey) galactic latitude samples with two magnitude limits.  Since this is a case where the results can be expected to depend sensitively on the assumptions we make about scale height, we include the corresponding histograms for model A2 and A3 (see Table~\ref{tab:models}) in Fig.~\ref{fig:gb}.  As expected, the ratio of long- to short-period systems is larger in the galactic plane samples in all cases; the effect is the smallest in model A2 (middle panels of Fig.~\ref{fig:gb}), because this is the model in which the galaxy comes the closest to being an isotropic distribution of stars.

\begin{figure}
 \includegraphics[width=84mm]{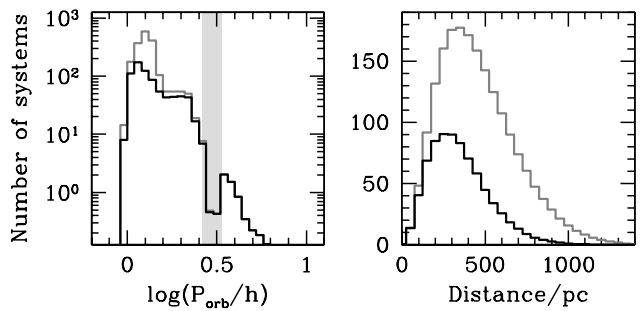}
 \caption{Period and distance distributions of all objects in the volume defined by $|b|>20^\circ$ (grey) and systems in that volume with $B<20$ (black).  Only 40\% of CVs in the volume limited sample are also in the magnitude limited sample, and practically all systems excluded by the magnitude limit have periods below the period gap, implying a strong bias against short-period systems.  Specifically, a $B<20$ survey is incapable of finding most of the systems near the period minimum.  This sample is from model A3, where the vertical density profile of stars is a Gaussian with a scale height of 190~pc.
}
 \label{fig:sdss}
\end{figure}

In a survey covering the area $|b|>20^\circ$, only 4.1\% of CVs in the survey volume are detected above a magnitude limit of $B<20$ in the case of model A1.  For model A3, 40\% of all systems in the survey volume have $B<20$, and less than 0.5\% of systems detected to this magnitude limit are long period systems; Fig.~\ref{fig:sdss} illustrates this.  The Sloan Digital Sky Survey (SDSS) reaches to $g=20$, and has sky coverage that can be roughly approximated as $|b|>20^\circ$ (e.g. \citealt{sdss}).  Because CVs found in the SDSS will in future comprise the largest uniformly selected sample, this survey is bound to be very important in the context of observational studies of CV populations (e.g. \citealt{sdsscvs}).  However, the result illustrated in Fig.~\ref{fig:sdss} implies that, even assuming a galactic scale height that is probably unrealistically small, the SDSS CV sample will by no means be volume limited.  Clearly, even if no colour bias existed in the SDSS CV sample, modelling of selection effects is required to interpret it correctly.  

More fundamentally, accepting that CV sub-populations with different typical ages do not share the same vertical galactic distribution, a volume limited, high-$|b|$ sample will in any case not reflect the intrinsic galactic population.  In fact, it is not practically possible to construct a volume limited observational sample in such a way that it represents the intrinsic population.  

\section{Sensitivity of the results to our assumptions}
Since there are considerable uncertainties regarding the DN outburst properties of CVs and their distribution in the Galaxy, we explore here the effect of varying our assumptions, by comparing several combinations of assumed galactic structure and outburst behaviour.  In addition, we consider whether our results are sensitive to the viscosity parameter we use in our disc SED models, since this is another potentially important source of error.

\begin{figure}
 \includegraphics[width=84mm]{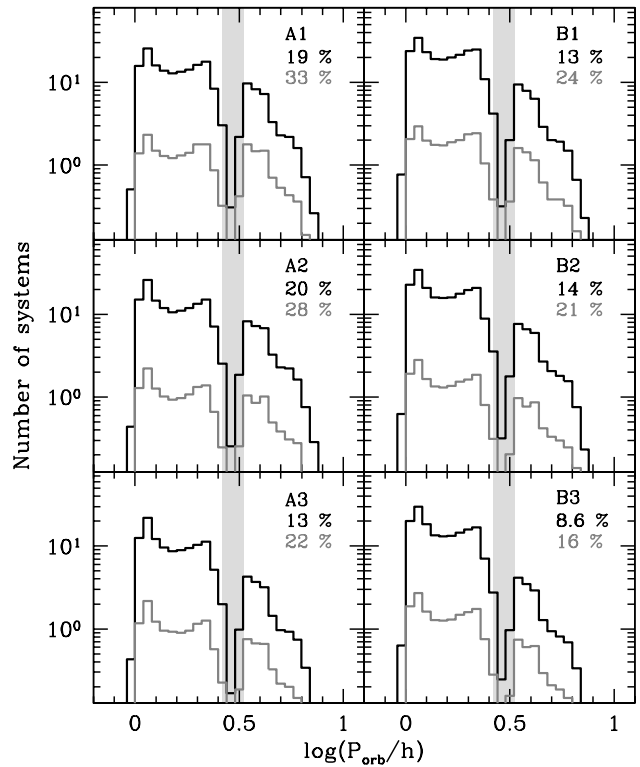}
 \caption{The orbital period distributions of magnitude limited samples produced by the first 6 models listed in Table~\ref{tab:models}.  These are in every case for all-sky samples.  Black (dark grey) histograms are for a magnitude limit of $V=16$ ($V=14$).  In each panel, the contribution of long-period systems is given as a percentage of the total for the fainter (black) and brighter (dark grey) magnitude limits.
}
 \label{fig:models}
\end{figure}

At fainter magnitude limits, all our predicted magnitude limited samples become more like the intrinsic distribution, and hence more alike.  Therefore we consider here samples with bright limiting magnitudes.  Fig.~\ref{fig:models} displays period histograms of magnitude limited samples produced by the first six models listed in Table~\ref{tab:models}.  The grey histograms are samples with $V<14$, while black shows $V<16$; the fraction of the total made up by long-period systems is given in each panel for both magnitude limits.  This fraction does not depend very sensitively on the assumptions about galactic structure and DN outburst properties.  As expected, it is more difficult to hide short-period systems if all CVs are assumed to accrete permanently at their long-term average rate, regardless of the adopted galaxy model.  The fraction of long-period systems increases towards brighter magnitude limits in all cases.

\begin{figure}
 \includegraphics[width=84mm]{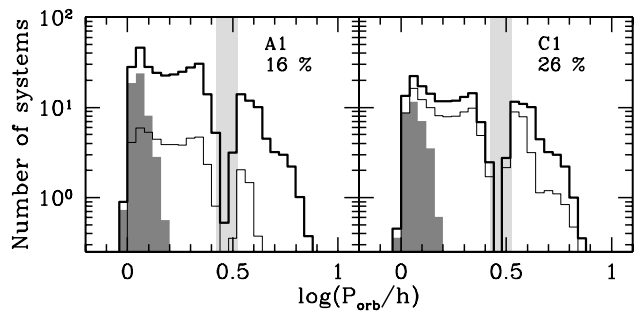}
 \caption{The orbital period distributions of samples produced by model A1 (left) and C1 (right).  Both samples have a magnitude limit of $V<16.5$.  Dark grey shading and fine black histograms show the contributions of period bouncers and DN in outburst respectively.  The contribution of long-period systems is given as a percentage of the total in each panel.
}
 \label{fig:a1c1}
\end{figure}

Model C1 produces the largest ratios of long- to short-period systems, and differs from, e.g., model A1 also in predicting fewer systems at a given magnitude limit, and in that relatively more detected systems are DN in outburst.  Fig.~\ref{fig:a1c1} compares the period distributions of all-sky samples with $V<16.5$ resulting from model A1 and C1.

Because of the Balmer discontinuity, the $U$ band is the region of the optical spectrum where the SED modelling of the disc emission is the least reliable.  Figure 15 of \cite{disc} shows that the $U-B$ colour of low-$\dot{M}$ discs is sensitive to the value of $\alpha$ (this sensitivity disappears at high $\dot{M}$).  For fixed inclination, varying $\alpha$ form 1 to 0.1 can change $U-B$ by almost 0.35~mag for $\dot{M}_d = 10^{14}\,\mathrm{g/s}$ and $r_d=10^{10}\,\mathrm{cm}$ (the effect is larger for larger $r_d$, but the low-$\dot{M}$ discs in our model have small outer radii).  In the very low-$\dot{M}$ systems where we take $\alpha=0.01$, the disc radiation has a negligible effect on the magnitudes and colours.  Anticipating the application in Section~\ref{sec:pg}, we are for the moment only concerned with the influence that incorrect $U-B$ colours may have when $U-B$ cuts are applied to samples with bright magnitude limits.  To determine whether an error of $\simeq0.35$~mag in the disc $U-B$ in systems with  $10^{13}\,\mathrm{g/s} \la \dot{M}_d \la 10^{14}\,\mathrm{g/s}$ is important, we performed the following test.  The $U-B$ colour of the disc is changed by adding 0.4~mag to the intrinsic disc $U$ magnitude for all systems with $\dot{M}_d < 2 \times 10^{14}\,\mathrm{g/s}$.  Taking the $B<16.2$ sample produced by model A1, less than 0.5\% of systems are lost from samples with blue cuts of $U-B < -0.7$ and $U-B < -0.9$ when their original colours are altered in this way.  The reasons are that low-$\dot{M}$ systems are rare in the magnitude limited sample, and that the flux contribution of the disc is increasingly unimportant compared to that of the white dwarf at lower accretion rates.

\section{Application to the Palomar-Green CV sample}
\label{sec:pg}
The CVs collected in the \cite{rkcat} catalogue include systems that were discovered on the basis of very different characteristics, such as physically distinct types of large amplitude variability, X-ray emission, blue colours, and emission lines.  The resulting overall sample is too heterogeneous to permit a reliable evaluation of its selection biases.  Note also that, in addition to the discovery bias, a second bias is introduced in the \cite{rkcat} sample by the need to find orbital periods.  This is magnitude dependent, but it is also, e.g., easier to measure shorter periods (although this is not very important at $P_{orb} \la 8$~h) and there is a bias towards magnetic systems in deciding which CVs are worthy of follow-up observations to find $P_{orb}$.  A quantitative comparison between theory and observations requires a reasonably large observed sample that has well defined selection criteria and quantified completeness.  At the moment the observational sample that conforms to these requirements the most closely is the sample of CVs found in the PG Survey.  

The PG Survey covered 10\,700~deg$^2$ at $|b|>30^\circ$ and was intended to have a magnitude limit of $B<16.16$ and a colour selection criterion of $U-B<-0.46$ \citep{pg}.  The blue cut and magnitude limit were always known to vary from field to field; the error in $U-B$ and the completeness were thought to be 0.39~mag and 84\% respectively.  More recent analyses of the PG data have shown that the actual magnitude limit and blue cut used by the survey were $B<16.16$ and $U-B<-0.71$, with errors of $\sigma_B=0.34$ and $\sigma_{U-B}=0.24$ (\citealt{pg_3}; see also \citealt{pg_wd}).  The survey produced a sample of 36 CVs, of which 3 are magnetic systems, and 6 have $P_{orb}>5\,\mathrm{h}$ (\citealt{ringwald}; Ringwald, private communication).  Depending on the nature of V378 Peg (the only PG CV that still does not have a period measurement) the remaining 27 systems form a statistically complete sample of unevolved, non-magnetic CVs.  

\begin{figure}
 \includegraphics[width=84mm]{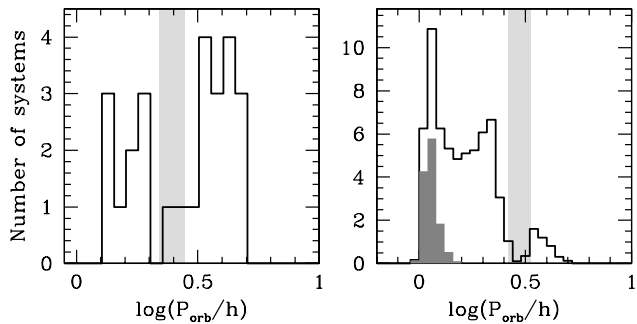}
 \caption{The orbital period distributions of the PG CV sample (including only non-magnetic system at $P_{orb} < 5\,\mathrm{h}$; left hand panel), and the predicted sample (also with systems at $P_{orb} \ge 5\,\mathrm{h}$ omitted; right hand panel).  Light grey shading of the range $2.2\,\mathrm{h} < P_{orb} < 2.8\,\mathrm{h}$ in the left hand panel, and $2.63\,\mathrm{h} < P_{orb} < 3.35\,\mathrm{h}$ in the right hand panel indicate the observed and model period gaps.  Dark grey shading represents period bouncers in the predicted sample.
}
 \label{fig:pg}
\end{figure}

Following \cite{pg_3}, we take the probability of a system with $B=B'$ and $U-B=(U-B)'$ being detected in the PG Survey as $P_1P_2$, where
{\scriptsize
\begin{displaymath}
P_1=1 - \frac{1}{\sigma_B\sqrt{2\pi}}\int_{-\infty}^{B'}\exp \left [ -\frac{1}{2} \left ( \frac{B-16.16}{\sigma_B} \right )^2 \right ] dB
\end{displaymath}
}
\noindent 
and
{\scriptsize
\begin{displaymath}
P_2=1 - \frac{1}{\sigma_{U-B}\sqrt{2\pi}}\int_{-\infty}^{(U-B)'}\exp \left [ -\frac{1}{2} \left ( \frac{(U-B)+0.71}{\sigma_{U-B}} \right )^2 \right ] d(U-B).
\end{displaymath}
}
\noindent
We applied this detection efficiency to systems at $|b|>30^\circ$ and $P_{orb}<5$~h in the population produced by our standard model A1.  A period histogram of the resulting predicted PG CV population is shown in Fig.~\ref{fig:pg}, together with the distribution of the observed sample.  

It is evident from Fig.~\ref{fig:pg} that the distribution produced by the simulation is inconsistent with the observed one, the most obvious discrepancy being that the ratio of long- to short-period CVs is much higher in the observed sample than is predicted.  If we conservatively assume that V378 Peg is a short-period system, and taking the upper edge of the period gap as 3~h, 14 out of 27 observed systems (52\%) at $P_{orb}<5$~h have periods above the gap.  The prediction is only 6.7\% (i.e. 2 systems). According to the binomial distribution, the probability that these numbers are consistent is less than $3 \times 10^{-10}$.  We also note that none of the PG CVs is considered to be a candidate period bouncer, whereas the simulation predicts that 19\% of the sample (i.e. 5 systems) should be period bouncers.

It was shown in Section 4 that the properties of the predicted detected sample are fairly insensitive to poorly constrained aspects of our modelling of disc SEDs, galactic structure, and DN outburst properties.  Of all the models we have considered, model C1 allows for the largest population of short-period systems to remain undetected, because it combines a large galactic scale height for these systems with an extreme assumption of DN outburst behaviour.  Repeating this simulation using model C1 yields a prediction of 10\% long-period systems in the PG sample, which is still inconsistent with the observed fraction.  Based on this and also on the sheer size of the discrepancy between the predicted and observed PG samples, we are confident that the predicted intrinsic population that is the basis of our modelling is ruled out by observations.  We believe this is the strongest evidence yet that the standard disrupted magnetic braking model for CV evolution, with strong magnetic braking above the period gap and none below the gap, is incorrect.  As has long been suspected, it predicts too many short-period systems, and, specifically, too many period bouncers.

\section{Conclusions}
We have presented a Monte-Carlo technique that allows us to model selection effects in observed samples of CVs.  Given any theoretical intrinsic population and any survey with well defined selection criteria, this technique can be applied to predict the observed population.  We have explored the impact of flux limits, colour cuts, and observing restricted ranges of galactic latitude, and finally applied our method to the CV sample produced by the PG Survey.  Our main results are the following.
\begin{enumerate}
\item The mass-transfer rates predicted by standard CV evolution theory are inconsistent with empirical estimates of the absolute magnitudes of DN in outburst, unless outburst duty cycles are much lower than is generally assumed (i.e. $<10$\% for normal DN and $<0.5$\% for period bouncers).
\item For practically achievable flux limits, no flux limited sample can be expected to reproduce the intrinsic CV population.  It is therefore quite inappropriate to compare an observed sample with the predicted intrinsic population. 
\item The properties of the period distribution of a magnitude limited sample depend strongly on the magnitude limit.  The ratio of short- to long-period systems and the prominence of the spike near the period minimum decrease at brighter magnitude limits.
\item The apparent magnitude distribution of CVs suggests that the known CV sample is not approximately magnitude-limited, even if a limiting magnitude as bright as $V=13$ is adopted.
\item A blue selection cut of $U-B\la-0.7$ introduces a serious bias against detecting short-period systems.  However, for existing $UV$-excess surveys, this bias is unimportant compared to the effect of the survey flux limits. 
\item Magnitude limited surveys at high galactic latitude are expected to yield samples that are very different from those produced by galactic plane surveys, because they detect a larger fraction of all systems inside the volume defined by the galactic latitude range.  However, a survey such as the SDSS is still not deep enough to be volume limited for CVs.  This means that, for some time to come, the effect of a magnitude limit will have to be considered when comparing an observed sample to theory.
\item A simulation of the selection effects present in the PG CV sample shows that observational biases are not sufficient to reconcile the intrinsic population predicted by standard CV evolution theory (with strong magnetic braking above the period gap) with the observed PG sample.  The real intrinsic CV population cannot contain as large a fraction of short-period systems (and, specifically, period bouncers) as is predicted by theory.
\end{enumerate}

\section*{Acknowledgements}
MLP acknowledges financial support from the South African National Research Foundation and the University of Southampton.  We thank Romuald Tylenda for the use of his accretion disc model, and Brian Warner for helpful discussion.  Fred Ringwald kindly provided an updated list of CVs discovered in the PG survey.

\bsp

\label{lastpage}

\end{document}